\documentclass[10pt, conference, letterpaper]{IEEEtran}
\IEEEoverridecommandlockouts
\usepackage{cite}
\usepackage{amsmath,amssymb,amsfonts}
\usepackage{algorithmic}
\usepackage{graphicx}
\usepackage{textcomp}
\usepackage[table]{xcolor}
\usepackage{tcolorbox}
\usepackage{colortbl}
\usepackage{multirow}
\usepackage{makecell}
\usepackage{subfig}
\usepackage{tcolorbox}
\usepackage{color}
\usepackage{enumitem}
\usepackage[flushleft]{threeparttable}
\allowdisplaybreaks[4]

\def\BibTeX{{\rm B\kern-.05em{\sc i\kern-.025em b}\kern-.08em
    T\kern-.1667em\lower.7ex\hbox{E}\kern-.125emX}}
\usepackage[pagebackref=false,breaklinks=true,colorlinks, bookmarks=false]{hyperref}
\newlength\savedwidth

\begin{document}
\addtolength{\textheight}{-0.01in}
\title{Reasoning AI Performance Degradation in 6G Networks with Large Language Models

}

\author{Liming Huang, Yulei Wu, Dimitra Simeonidou\\
	\IEEEauthorblockA{School of Electrical, Electronic and Mechanical Engineering, University of Bristol, Bristol, U.K.\\
		liming.huang@bristol.ac.uk, y.l.wu@bristol.ac.uk, dimitra.simeonidou@bristol.ac.uk}
}

\maketitle

\begin{abstract}
The integration of Artificial Intelligence (AI) within 6G networks is poised to revolutionize connectivity, reliability, and intelligent decision-making. However, the performance of AI models in these networks is crucial, as any decline can significantly impact network efficiency and the services it supports. Understanding the root causes of performance degradation is essential for maintaining optimal network functionality. In this paper, we propose a novel approach to reason about AI model performance degradation in 6G networks using the Large Language Models (LLMs) empowered Chain-of-Thought (CoT) method. Our approach employs an LLM as a ``teacher'' model through zero-shot prompting to generate teaching CoT rationales, followed by a CoT ``student'' model that is fine-tuned by the generated teaching data for learning to reason about performance declines. The efficacy of this model is evaluated in a real-world scenario involving a real-time 3D rendering task with multi-Access Technologies (mATs) including WiFi, 5G, and LiFi for data transmission. Experimental results show that our approach achieves over 97\% reasoning accuracy on the built test questions, confirming the validity of our collected dataset and the effectiveness of the LLM-CoT method. Our findings highlight the potential of LLMs in enhancing the reliability and efficiency of 6G networks, representing a significant advancement in the evolution of AI-native network infrastructures.

\end{abstract}


\section{Introduction}

As technology continues to evolve, network infrastructure is rapidly advancing towards the next-generation framework, notably 6G. Unlike its predecessors, 6G is expected to feature a deep and pervasive integration of Artificial Intelligence (AI)~\cite{9349624}. The native AI integration has the potential to significantly enhance the functionality and efficiency of 6G networks, offering powerful capabilities in terms of connectivity, reliability, and intelligent decision-making. However, the performance of AI models within these networks is critical. Any degradation in AI performance can significantly affect network operations and the services they support. Various factors can contribute to AI performance degradation, such as data quality issues, algorithmic biases, and sudden changes in network conditions. Moreover, the complex and dynamic nature of 6G can deteriorate these challenges, as AI models may struggle to adapt swiftly to the high-speed, voluminous data flows that characterize such advanced systems.

To address the challenges posed by AI performance degradation, numerous AI monitoring systems have been developed to detect declines in model performance~\cite{10207029,9978638}. These systems act as early warning mechanisms, identifying performance issues that could potentially compromise network efficiency and reliability. However, simply detecting a decline without understanding its root causes can be insufficient for effective intervention. This gap can lead to delays in problem resolution, resulting in prolonged periods of suboptimal network performance. By contrast, integrating reasoning capabilities within these systems allows for more effective diagnostic processes, enabling networks to implement corrective actions swiftly. Consequently, the efficiency and reliability of 6G networks can be further enhanced, allowing AI-native operations to realize their full potential with minimal disruptions.

Current approaches to reasoning about model performance degradation primarily focus on isolated AI models, without considering the broader dynamic network environment. For example, drift detection~\cite{9076305} is commonly employed to identify performance degradation at the data level, detecting input changes and highlighting new or emerging data patterns. In addition, some heuristic methods~\cite{ahn2022heuristics} use historical data to infer potential causes of performance issues. While these methods can be somewhat effective, they face significant challenges in the context of 6G networks. First, the massive volume and high velocity of data within future networks can overwhelm traditional analytics, making real-time processing difficult. Second, the complexity of 6G network architectures and AI models introduces a degree of variability and unpredictability that these approaches cannot handle. Third, current methods often require extensive domain knowledge and manual intervention to interpret data and draw inferences, which limits their scalability and adaptability. Thus, there is an urgent need for more autonomous and intelligent reasoning mechanisms capable of overcoming these challenges in 6G.

Large Language Models (LLMs) represent a revolutionary AI advancement, characterized by their ability to understand, generate, and interpret human language with remarkable accuracy~\cite{chang2023survey}. Powered by great amounts of data and advanced neural network architectures, LLMs can discern patterns and relationships within complex datasets, making them well-suited for tasks requiring deep understanding and reasoning. 

In this paper, we introduce an innovative LLM-powered Chain-of-Thought (CoT) method (LLM-CoT) for reasoning about AI performance degradation in 6G networks. The proposed LLM-CoT includes a ``teacher'' model, responsible for generating CoT rationales as training data, and a ``student'' model, which is a CoT model fine-tuned using this data. First, we employ a three-step zero-shot prompting to generate CoT rationales by a teacher LLM model, including network lecture generation, plan generation and rationale generation. The generated CoT rationales are used as teaching data that embody insights and reasoning aligned with network performance dynamics. Next, the CoT student model undergoes fine-tuning through a two-stage process. The first stage, ``rationale generation teaching,'' focuses on training the model to construct logical explanations for observed phenomena or data points. In the second stage, ``answer inference teaching,'' the model employs the previously generated rationales to infer answers to specific problems. During inference, the rationales produced in the first stage guide the model in making informed decisions based on test data, thereby enhancing its understanding and reasoning capabilities with respect to AI performance issues in 6G environments.

To evaluate the effectiveness of our LLM-CoT model, we deploy it in the context of a real-time 3D rendering task, a scenario that demands rapid data transmission and high-precision processing. In this setup, multiple cameras capture scene images, which are then transmitted to an edge AI model, a 3D Gaussian Splatting (3D-GS) model, for rendering the 3D scene. The images are transmitted using three wireless technologies: WiFi, 5G, and LiFi, to maximize throughput and minimize latency. As the 3D-GS model begins to produce progressively worse renderings, identifying the root causes of this performance degradation becomes critical. It is in this complex and dynamic environment that our LLM-CoT model proves its value. By reasoning through the intricacies of network performance, data quality, and transmission protocols, our model provides accurate insights into the factors affecting the edge AI model's output. Evaluations confirm our LLM-CoT model can achieve reliable accuracy in diagnosing AI performance degradation in 6G networks.

Our main contributions are summarized as follows:
\vspace{-2pt}
\begin{enumerate}
\item For the first time, we apply LLMs to the task of reasoning about AI model performance degradation within the context of native AI networks and 6G environments. This approach can be used for diagnosing and mitigating performance issues in the increasingly AI-native landscape of 6G networks, showcasing a novel application of LLMs beyond traditional use cases.
\item We propose an innovative LLM-powered CoT model designed for detailed reasoning tasks. Our method leverages zero-shot prompting of LLMs to generate rationale teaching data, which is then used to fine-tune a CoT student model, enabling it to perform sophisticated reasoning processes.
\item We implement and evaluate our LLM-CoT model in a real-world application scenario involving a real-time 3D rendering task. Through this task, we create a new dataset named AI-DR, consisting of 4,160 reasoning questions for evaluation. Experiments demonstrate that our LLM-CoT method achieves over 97\% reasoning accuracy, validating the effectiveness of our approach and confirming the dataset's utility and relevance.
\end{enumerate}

\section{Related Work}

\subsection{Reasoning AI Performance in Networks}

Recent research has increasingly emphasized the deployment of AI models within communication networks~\cite{10622263,karapantelakis2024generative}, highlighting the critical need for robust monitoring and reasoning frameworks. However, current efforts in reasoning AI performance primarily focus on the models in isolation, neglecting the dynamic and complex nature of the network environment in which these models operate. For example, Liu et al.~\cite{9076305} analyzed the concept drift of input data and proposed an equal intensity k-means approach for its detection. Ahn et al.~\cite{ahn2022heuristics} proposed a heuristics-based model to reason about performance issues with multiple causes and effects. 

While these approaches address drift detection and heuristic learning at the data and model design levels, they do not incorporate network analysis into their reasoning processes. This omission significantly lowers the effectiveness of root cause analysis, particularly in the context of modern and next-generation communication networks.

\subsection{LLMs for Intelligent Networks}

Large Language Models (LLMs), with their ability to process and generate human-like text, represent a significant advancement over traditional AI models due to their scale, versatility, and deep contextual understanding. Unlike earlier AI models that were often task-specific, LLMs can generalize across a wide range of tasks, making them particularly well-suited for the complex and dynamic nature of network systems.

In intelligent networks, LLMs have demonstrated potential in automating complex management tasks, such as enhancing security protocols~\cite{meng2024large}, improving predictive analytics for traffic management~\cite{manias2024towards}, and optimizing resource allocation~\cite{liu2024resource}. Furthermore, LLMs can interpret and automate network policy configurations, significantly reducing manual oversight and the potential for errors~\cite{li2024preconfig,bandara2024slicegpt}.

As AI-native 6G networks evolve, integrating LLMs into network infrastructures is poised to become a crucial component, driving the next wave of autonomous network systems. The scalability and adaptability of LLMs make them essential for advancing network intelligence and efficiency, leading to a new era of intelligent network evolution. In this paper, we propose an LLM-CoT model that significantly advances the AI reasoning for 6G. It leverages the depth and adaptability of LLMs to analyze AI performance and 6G networks, providing a promising solution to accurately identify the roots of performance degradation across networks and model design factors.

\section{Task Description and Problem Formulation}

\subsection{Edge Tasks: Real-Time 3D Rendering} 

\subsubsection{Task Description} \label{sec:edge-task}
\begin{figure}
	\centering
        \vspace{2pt}
	\includegraphics[width=1\linewidth]{"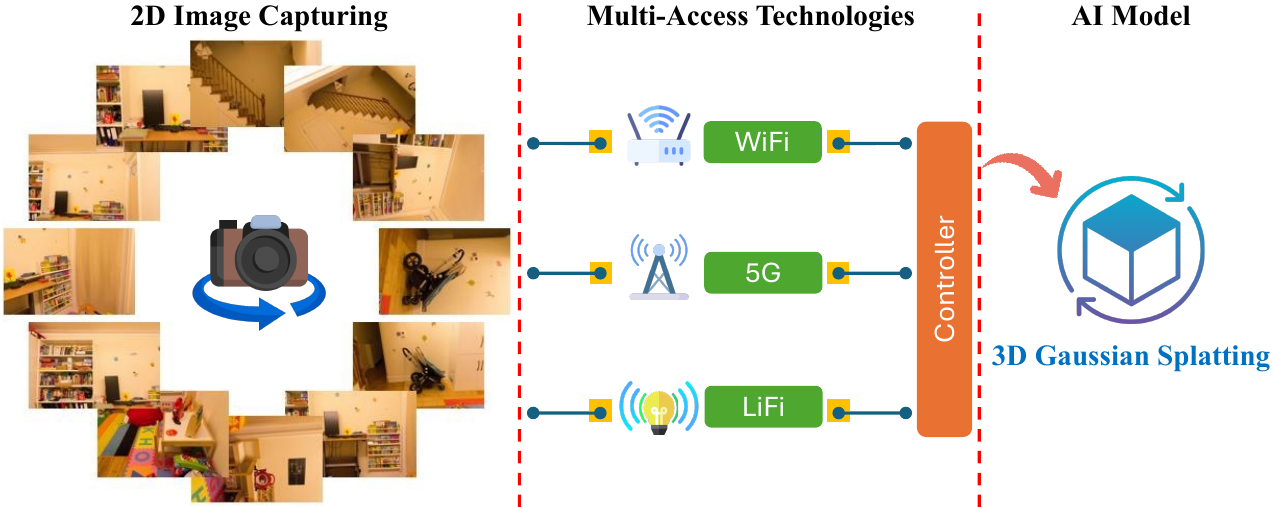"}	
	\caption{The 3D rendering using 3D-GS model under 6G mATs.}	
	\label{fig:3d-task}
        \vspace{-5pt}
\end{figure}

3D rendering is an application aimed at creating digital 3D models from 2D images captured in real environments. In this paper, we consider a real-time 3D rendering task within the 6G network environment. As shown in Fig.~\ref{fig:3d-task}, cameras capture 2D images in real-time and these images are transmitted through wireless networks into the edge AI model for completing the 3D rendering.

\subsubsection{Edge AI Models} \label{sec:edge-ai}
This paper leverages a 3D rendering model named 3D Gaussian Splatting (3D-GS)~\cite{kerbl20233d} for this task. The 3D-GS model is a real-time rendering model that can achieve great visual quality while allowing real-time ($\geq$ 100 fps) novel-view synthesis at 1080p resolution.

\subsection{6G Network Environment: mATs} \label{sec:m-at}

For real-time 3D rendering tasks, we use multi-Access Technologies (mATs) for data transmission. Specifically, the mATs encompass three wireless accesses including WiFi, 5G and LiFi, which can leverage the unique advantages of each access technology to ensure high throughput and low-latency transmission of volumetric data. The mATs emphasize the importance of high reliability and connectivity required to enable next-generation digital services and experiences in 6G.

\subsection{Problem Formulation for Reasoning}

This paper addresses the issue of reasoning AI performance degradation in complex 6G network environments. The analysis focuses on an edge computing task, 3D rendering, as discussed in Section~\ref{sec:edge-task}. The AI model used for this task is the 3D-GS model, as detailed in Section~\ref{sec:edge-ai}. The 6G network environment is characterized by its use of mATs-featured wireless connections, as outlined in Section~\ref{sec:m-at}.

\begin{figure*}
	\centering
        \vspace{2pt}
	\includegraphics[width=0.95\linewidth]{"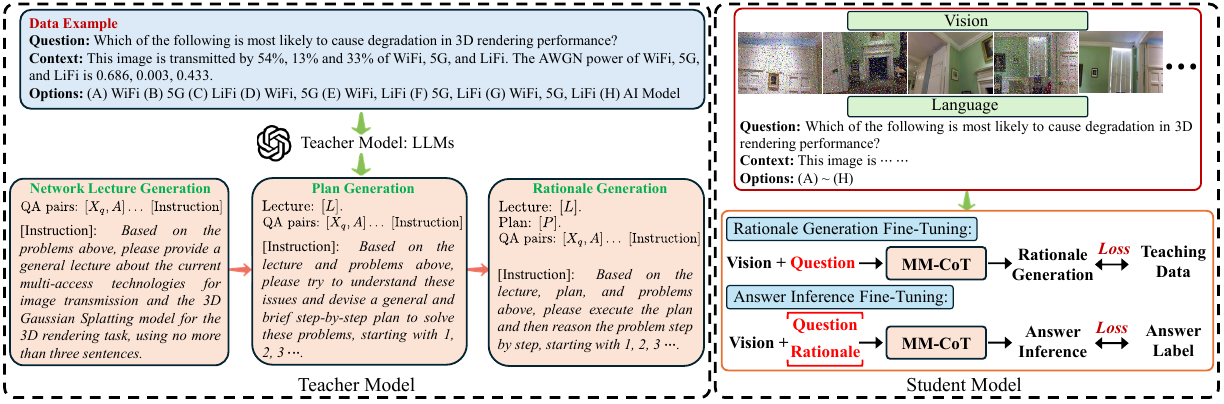"}	
	\caption{The framework of our LLM-CoT methodology for reasoning AI performance degradation in 6G networks.}	
	\label{fig:framework}
        \vspace{-5pt}
\end{figure*}

The reasoning framework employed is the Multimodal-CoT (MM-CoT)~\cite{zhang2023multimodal}. In this framework, the model denoted as $M$, processes multi-modal inputs: the language input $X_L$ and the vision input $X_V$. The language input $X_L$ defines the reasoning questions. The question format is formulated as: ``Question: Which of the following is most likely to cause degradation in 3D rendering performance?'' and ``Options: (A)-(G) mATs-based factors; (H) AI Model''. Furthermore, we include the context information about the mATs environment into the question as shown in Fig.~\ref{fig:framework}. The vision input $X_V$ comprises image features extracted from multi-view images that are used in the real-time 3D rendering. 

The output of the reasoning process is twofold: rationale generation and answer inference. The \textit{rationale} provides the step-by-step problem analysis, i.e., chain-of-thought. The \textit{answer inference} identifies the factors that lead to AI performance degradation, which may include network-related or internal model issues.

\section{LLM-Powered Chain-of-Thought Reasoning}

\subsection{Overview}

To accurately and comprehensively reason about AI performance degradation in 6G networks, we employ a fine-tuning strategy for the LLM-CoT model. Our approach is structured into two primary phases: generating teaching data with a teacher model and fine-tuning the CoT student model.

In the first phase, we leverage the advanced LLM GPT-3.5 as the teacher model to generate teaching rationales through zero-shot prompting. In this stage, we adopt a Plan-based CoT (PCoT) strategy~\cite{wang2023plan} for prompting that encompasses network lecture generation, plan generation, and rationale generation. This methodology enables the LLM to deeply understand edge tasks, edge AI models, and 6G network environments, thereby enhancing its rationale generation capabilities.

In the second phase, the rationales generated by the teacher model are used to fine-tune the student model. In this paper, we adopt the MM-CoT model~\cite{zhang2023multimodal} as the student model. A two-stage fine-tuning framework is applied to the MM-CoT model for rationale generation and answer inference teaching. Through this approach, the student model can accurately identify the causes of AI performance degradation and provide a logical CoT explanation, achieving comprehensive reasoning.

\subsection{Teacher Model}

Given the complexity of 6G networks and their extensive use of AI, relying on human-annotated CoT signals is impractical. Manual annotation is not only time-consuming but also demands high-level expertise from human annotators. To address these challenges, this paper adopts LLMs to generate CoT rationales via zero-shot prompting. Specifically, we employ LLM GPT-3.5 as the teacher model for generating teaching data. While GPT-3.5 possesses extensive knowledge across various fields, fully and accurately understanding AI operations within complex 6G network environments remains a significant challenge. Consequently, we adopt a zero-shot plan-and-solve prompting strategy~\cite{wang2023plan} to generate PCoT. This approach innovatively designs a three-step zero-shot prompting for PCoT generation, tailored to 6G mATs and the edge AI model for 3D rendering.

\textbf{Step 1: Network Lecture Generation:}
The first step involves prompting GPT-3.5 to generate comprehensive network lectures covering essential knowledge about the 6G mATs, and the specifications of the 3D rendering AI model and task. We use the following template for the lecture prompt: ``QA pairs: $[X_q, A] \dots [\textit{Instruction}]$.'' The $[\textit{Instruction}]$ is formulated as: ``\textit{Based on the problems above, please provide a general lecture about the current multi-access technologies for image transmission and the 3D Gaussian Splatting model for the 3D rendering task, using no more than three sentences.}''

\textbf{Step 2: Plan Generation:}
With the foundational lecture established, the second step prompts the teacher model to create a step-by-step plan for addressing the 3D rendering task within 6G environments. The plan generation template is: ``Lecture: [$L$]. QA pairs: $[X_q, A] \dots [\textit{Instruction}]$.'' The $[\textit{Instruction}]$ is: ``\textit{Based on the lecture and problems above, please try to understand these issues and devise a general and brief step-by-step plan to solve these problems, starting with $1, 2, 3 \dots$.}''

\textbf{Step 3: Rationale Generation:}
The final step combines the lecture and plan to generate PCoT rationales concerning AI model performance degradation, which are then used for training each example. The rationale generation template is: ``Lecture: [$L$]. Plan: [$P$]. QA pairs: $[X_q, A] \dots [\textit{Instruction}]$.'' The $[\textit{Instruction}]$ is: ``\textit{Based on the lecture, plan, and problems above, please execute the plan and then reason the problem step by step, starting with $1, 2, 3 \dots$.}''

Through these three steps, we generate PCoT rationales for reasoning about AI model performance degradation within the complex 6G environments. Equipped with these PCoT rationales, the student model is trained to generate informed rationales and make intelligent inferences about network performance and AI model behaviors.

\subsection{Student Model}

The student model employs a MM-CoT framework~\cite{zhang2023multimodal}. It processes both language and image inputs. The learning process involves a two-stage fine-tuning approach, which includes training for rationale generation and answer inference.

\subsubsection{\textbf{MM-CoT Architecture}} \label{sec:mmcot}

The MM-CoT architecture comprises three main components: encoding, fusion, and decoding. In the encoding stage, the language input $X_L$ and the vision input $X_V$ are processed using a language Transformer and a Vision Transformer (ViT~\cite{dosovitskiy2021an}) respectively:
\begin{align} \label{eq:encoding}
		&H_L = \text{L-Encoder}(X_L), \\
		&H_V = \text{V-Encoder}(X_V),
\end{align}
where $\text{L-Encoder}$ and $\text{V-Encoder}$ are the language and vision encoders, and $H_L$ and $H_V$ are the resulting text representations and image features.

For information fusion, the text representation $H_L$ and the image feature $H_V$ are combined. First, a single-head attention network~\cite{zhang2023multimodal} is used to transform the image patches to align with the text tokens. The transformed image feature is denoted as $H_V^a$. Then, following~\cite{zhang2019neural}, the language and vision information fusion is performed as:
\begin{equation}
	H_f = (1-\lambda) H_L + \lambda H_V^a,
\end{equation}
where $H_f$ is the fused output, $\lambda$ is a gating mechanism calculated by $\lambda = \text{Sigmoid} (W_L H_L + W_V H_V^a)$ where $\text{Sigmoid}(\cdot)$ is the sigmoid function, and $W_L$ and $W_V$ are parameters.

The decoding process uses this fused information $H_f$ to predict the target through a Transformer decoder, $\text{L-Decoder}$:
\begin{equation}
    Y = \text{L-Decoder}(H_f).
\end{equation}

\subsubsection{\textbf{Rationale Generation Fine-Tuning}} 

In this stage, the MM-CoT model detailed in Section~\ref{sec:mmcot} is used for fine-tuning rationale generation. The language input is specified as $X_L^1$ and the vision input as $X_V$. The probability of generating a rationale text $Y_R$ of length $N_R$ is defined as:
\begin{equation}
    p(Y_R \mid X_L^1, X_V) = \prod_{i=1}^{N_R} p_{\theta_R} ((Y_R)_i \mid X_L^1, X_V, (Y_R)_{< i}),
\end{equation}
where $p_{\theta_R}(\cdot)$ denotes the rationale generation process within the MM-CoT framework, and $\theta_R$ represents the set of learnable parameters.

\begin{table*}[!t]
	\renewcommand{\arraystretch}{1.12}
	\setlength\tabcolsep{3.0pt}
        \centering
        \vspace{5pt}
        \caption{\small Main Results (\%) of Answer Inference And Rationale Generation Evaluation on The AI-DR Test Set}
        \vspace{-5pt}
	\label{table:main-result}
        \begin{threeparttable}
        \begin{tabular}{c|c|c|c|c|c|ccc}
        \hline
        \multicolumn{2}{c|}{\multirow{2}{*}{Model}} & \multirow{2}{*}{Year} & \multirow{2}{*}{Learning Mode} & \multirow{2}{*}{Size} & Answer & \multicolumn{3}{c}{Rationale} \\
        \cline{6-9}\multicolumn{2}{c|}{} &       &       &       & A-Acc     & BLEU-1  & ROUGE-L & Similarity \\
        \hline
        \multirow{7}{*}{\rotatebox{90}{Data Splitting I}} & GPT-3.5  & 2022     & zero-shot-prompting & $>$175B    & 75.96  & --- & --- & ---  \\
        & GPT-4  & 2024      & zero-shot-prompting & $>$175B    & 80.28  & --- & --- & ---  \\
        \cline{2-9}      & UnifiedQA-CoT$_\text{Base}$~\cite{lu2022learn}   & 2022      & fine-tuning      & 223M      & 92.78      & 71.41      & 78.77       & 94.52 \\
        & Multimodal-CoT$_\text{Base}$~\cite{zhang2023multimodal} & 2023      & fine-tuning      & 223M      & 95.91      & 73.96      & 81.02       & 94.91 \\
        & LLM-CoT$_\text{Base}$ (Ours) & --      & fine-tuning      & 223M      & 98.43      & 76.68      & 82.49      & 95.19 \\
        \cline{2-9} & Multimodal-CoT$_\text{Large}$~\cite{zhang2023multimodal} & 2023      & fine-tuning      & 738M      & 98.55      & 77.51      & 83.20      & 95.21 \\
        & LLM-CoT$_\text{Large}$ (Ours) & --      & fine-tuning      & 738M      & 99.27      & 78.68      & 84.49      & 96.19 \\
        \hline
        \hline
        \multirow{7}{*}{\rotatebox{90}{Data Splitting II}} & GPT-3.5 & 2022      & zero-shot-prompting      & $>$175B      & 71.61  & --- & --- & --- \\
        & GPT-4  & 2024      & zero-shot-prompting      & $>$175B      & 76.59  & --- & --- & ---   \\
        \cline{2-9}      & UnifiedQA-CoT$_\text{Base}$~\cite{lu2022learn}   & 2022      & fine-tuning      & 223M      &  90.83     & 69.17           & 75.71      & 92.81 \\
        & Multimodal-CoT$_\text{Base}$~\cite{zhang2023multimodal} & 2023      & fine-tuning      & 223M      & 93.92      & 72.79         & 79.56      & 94.33 \\
        & LLM-CoT$_\text{Base}$ (Ours) & --      & fine-tuning      & 223M      & 97.01      & 74.10       & 80.60      & 95.01\\
        \cline{2-9}      & Multimodal-CoT$_\text{large}$~\cite{zhang2023multimodal} & 2023      & fine-tuning      & 738M      & 97.11      & 74.49     & 80.87      & 94.88 \\
        & LLM-CoT$_\text{Large}$ (Ours) & --      & fine-tuning      & 738M      & 98.20      & 75.31       & 81.59      & 95.24 \\
        \hline
        \end{tabular}        
        \begin{tablenotes}
            \item {\scriptsize Data Splitting I: Randomly generating training, validation, and test sets in a 3:1:1 ratio.}
            \item {\scriptsize Data Splitting II: `Playroom' and `DrJohnson' as the training set, `Train' as the validation set, and `Truck' as the test set.} 
        \end{tablenotes}   
        \end{threeparttable}
        \vspace{-10pt}
\end{table*}

\begin{figure*}
	\centering
	\includegraphics[width=0.92\linewidth]{"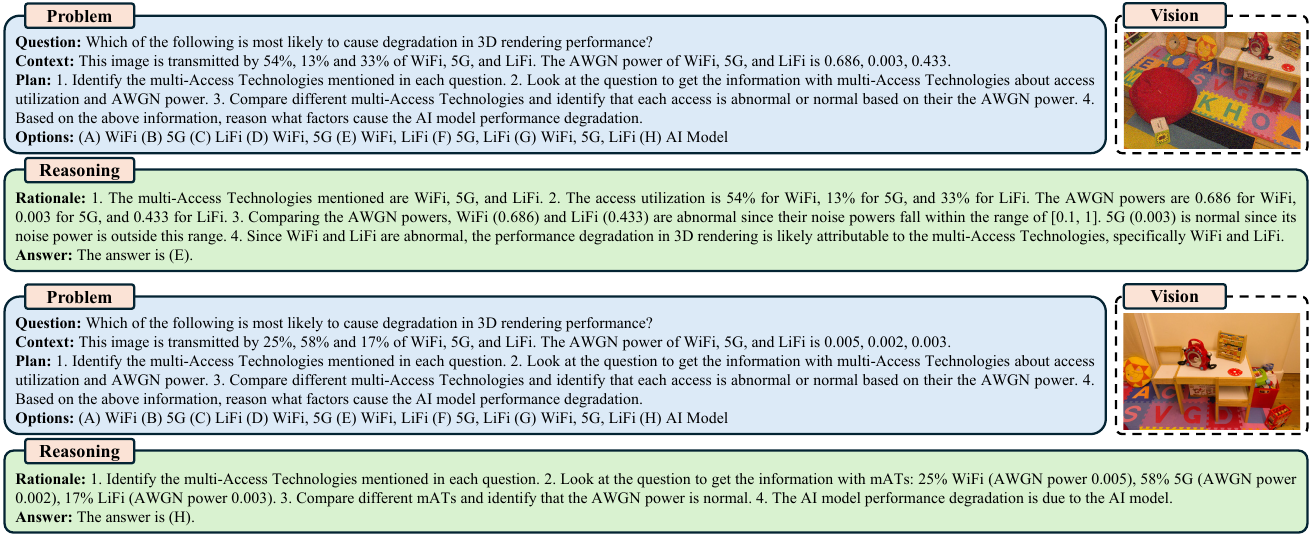"}	
	\caption{\small The examples of our LLM-CoT methodology for reasoning AI performance degradation in 6G networks.}	
	\label{fig:cot-example}
        \vspace{-10pt}
\end{figure*}

\subsubsection{\textbf{Answer Inference Fine-Tuning}} 

Following rationale generation, the language input from the first stage $X_L^1$ and the rationale output $Y_R$ are concatenated to form the second-stage language input $X_L^2$. Then, $X_L^2$ and $X_V$ are processed using the MM-CoT to predict the answer $Y_A$ of length $N_A$:
\begin{equation}
    p(Y_A \mid X_L^2, X_V) = \prod_{i=1}^{N_A} p_{\theta_A} ((Y_A)_i \mid X_L^2, X_V, (Y_A)_{< i}),
\end{equation}
where $p_{\theta_A}(\cdot)$ indicates the answer inference process, and $\theta_A$ is the corresponding set of learnable parameters.

\section{Experiment}

\subsection{Experimental Setup}

\textbf{Edge Task Simulation.} To capture real-time images, we adopt four datasets for mATs-based image transmission: ``Playroom'' (900 images) and ``DrJohnson'' (1052 images) from the Deep Blending dataset~\cite{hedman2018deep}, and ``Truck'' (1004 images) and ``Train'' (1204 images) from the Tank\&Temples dataset~\cite{kerbl20233d}. We employ three access technologies for transmission-WiFi, 5G, and LiFi-with average throughputs of 800 Mbps, 400 Mbps, and 200 Mbps, respectively. Images are received from cameras and converted to bits, which are then processed using Quadrature Frequency Shift Keying for modulation and demodulation to achieve RGB image transmissions. These channels are randomly subject to Additive White Gaussian Noise, which can distort the signal and impact the quality of transmitted images, thereby affecting the 3D rendering outcomes of the edge AI model 3D-GS.

\textbf{Dataset.} Given the absence of existing datasets for AI performance degradation in 6G, we develop a new dataset named AI Degradation Reasoning (AI-DR) to implement evaluations. This dataset follows the structure of the Science Question Answering (ScienceQA) dataset~\cite{lu2022learn}, a widely-used multi-modal reasoning dataset including extensive language-vision science questions. Following the format of ScienceQA~\cite{lu2022learn}, we create 4160 reasoning questions, each accompanied by an image and question data example. In this study, we employ two strategies for data splitting: I) The dataset is randomly divided into training, validation, and test sets in a 3:1:1 ratio. II) Specific subsets of the data are designated for each set to assess reasoning abilities across datasets, with `Playroom' and `DrJohnson' serving as the training set, `Train' as the validation set, and `Truck' as the test set. 

\textbf{Evaluation Metrics.} Following the ScienceQA~\cite{lu2022learn} and other reasoning studies~\cite{zhang2023multimodal, wang2024t}, we evaluate \textit{answer inference} performance using answer accuracy (A-Acc), which is the ratio of correct answers. For \textit{rationale generation} performance, we adopt Natural Language Processing (NLP) evaluation metrics including BLEU-1 (Bilingual Evaluation Understudy) Score, ROUGE-L (Recall-Oriented Understudy for Gisting Evaluation) Score, and the Similarity score. Detailed explanations of these metrics can be found in~\cite{pandey2023natural}. Overall, the accuracy of answer inference indicates the model's ability to accurately identify the factors leading to AI performance degradation. Meanwhile, the accuracy of the rationales assesses whether the model's CoT is precise.

\textbf{Baselines.} First, we include API-based OpenAI LLMs (GPT-3.5 and GPT-4) for comparison, which use zero-shot prompting for answer generation without fine-tuning. The prompting strategies of the two OpenAI LLMs follow the settings of~\cite{lu2022learn,lu2024chameleon}. Next, two state-of-the-art (SoTA) fine-tuning based CoT models are used for comparison, including UnifiedQA-CoT~\cite{lu2022learn} and Multimodal-CoT~\cite{zhang2023multimodal}.

\textbf{Implementation Details.} For rationale generation, the GPT-3.5-turbo model serves as the teacher. For the student model, we utilize both the Multimodal-CoT-T5-base with 223M parameters and the Multimodal-CoT-T5-large with 738M parameters for fine-tuning. The fine-tuning process consists of 30 epochs with a learning rate of 5e-5 and a maximum input length of 512 tokens. This project will be released at https://github.com/lmhuang-me/LLM-CoT-6G.git.

\subsection{Model Performance}

The main results of the answer and rationale generation evaluation are presented in Table~\ref{table:main-result}. For prompting models GPT-3.5 and GPT-4, we only evaluate their answer accuracy as comparing their un-fine-tuned rationale generation would be unfair. Table~\ref{table:main-result} shows that our LLM-CoT$_\text{Base}$ model outperforms the SoTA CoT fine-tuning models, UnifiedQA-CoT$_\text{Base}$ and Multimodal-CoT$_\text{Base}$, in answer accuracy (A-Acc). The improvements are 5.65\% and 2.52\% for Data Splitting I, and 6.18\% and 3.09\% for Data Splitting II, respectively. In the area of rationale generation, our LLM-CoT$_\text{Large}$ model demonstrates competitive performance, achieving scores of 78.68\%, 84.49\%, and 96.19\% for BLEU-1, ROUGE-L, and Similarity metrics respectively on Data Splitting I. Overall, the results confirm two main aspects: \textbf{1)} The SoTA models can effectively use our dataset to produce reasonable outcomes, thereby demonstrating the usability and validity of our dataset. \textbf{2)} Our model achieves competitive results, which proves the effectiveness of the solution we have designed. The example of our CoT reasoning is shown in Fig.~\ref{fig:cot-example}.

\section{Conclusion}

In this study, we pioneered the investigation into the performance degradation of AI in 6G networks using LLMs. Unlike previous efforts that focused solely on AI model structures, our approach considered the AI model within a 6G-enabled mATs network environment, taking into account the effects of various access technologies for reasoning analytics. We selected 3D rendering as the edge AI task with mATs for data transmission and introduced the LLM-CoT method for in-depth reasoning. A teacher model was developed to generate teaching data, and a student model was subsequently fine-tuned with this data for improved rationale and answer prediction. Furthermore, we created a dataset, named AI-DR, specifically for reasoning about AI performance degradation in 6G contexts. Experimental results confirmed the validity of our dataset and the effectiveness of the LLM-CoT method.

\section*{Acknowledgements}
{ This work has been partially sponsored by the UK GOV DSIT (FONRC) project REASON.}

\bibliographystyle{IEEEtran}
\bibliography{reference}

\end{document}